\documentclass[
    ,final            
  ]
  {aipproc}

\layoutstyle{6x9}

\begin{document}

\title{Quantum effects in the diffusion process to form a heavy nucleus
in heavy-ion fusion reactions}


\author{Kouhei Washiyama}{
  address={Department of Physics, Tohoku University, 
980-8578 Sendai, Japan}
}

\author{B\"ulent Yilmaz}{
  address={Physics Department, Ankara University, 
TR-06100 Ankara, Turkey},
altaddress={Physics Department, Middle East Technical University, 
TR-06531 Ankara, Turkey}
}

\author{Sakir Ayik}{
  address={Physics Department, Tennessee Technological 
University, Cookeville, Tennessee 38505, USA}
}

\author{Noboru~Takigawa}{
  address={Department of Physics, Tohoku University, 
980-8578 Sendai, Japan}
}

\begin{abstract}
We discuss quantum effects in the diffusion process which is used 
to describe the shape evolution from the touching configuration 
of fusing two nuclei to a compound nucleus. 
Applying the theory with quantum effects to the case where 
the potential field, the mass and friction 
parameters are adapted to realistic values of 
heavy-ion collisions, we show that 
the quantum effects play significant roles 
at low temperatures which are relevant to the synthesis 
of superheavy elements.

\end{abstract}

\maketitle


\section{Introduction}

It is now well accepted that 
it is not sufficient for the two nuclei in heavy-ion collisions 
to overcome the Coulomb barrier 
to form a heavy compound nucleus such as superheavy elements.
This is because the conditional saddle, which should be 
overcome for two nuclei to fuse, is located 
inside the Coulomb barrier for collisions between two heavy nuclei. 
This provides an origin of the so-called fusion hindrance 
phenomena \cite{swiatecki}. 
We thus need to describe the shape evolution 
from the touching configuration of fusing two nuclei to 
a more compact spherical-like compound nucleus 
by overcoming a potential barrier near the conditional saddle point.

A diffusion model has been applied
to describing this process,
especially to describing the formation of 
superheavy elements \cite{aritomo99,abe02,aritomo04}.
In these studies, so far 
the standard fluctuation-dissipation relation which holds 
at high temperatures has been postulated
to relate the diffusion coefficients 
to the friction coefficients.
Although these studies provide some illuminating information  
and look to be successful to some extent in the data analysis, 
one needs to carefully examine the validity of 
the standard fluctuation-dissipation relation 
in order to apply to the diffusion process at low temperatures
which are relevant to the synthesis of superheavy elements. 
Since superheavy elements are stabilized 
by shell correction energies, one has to 
synthesize them at reasonably low excitation energies, that is,
at low temperatures as low as 1 MeV or below. 
On the other hand, the barrier curvature around 
the conditional saddle point
is also of the order 
of 1 MeV. It is thus likely that quantum effects
play an important role in the compound nucleus formation process,
especially in the synthesis of superheavy elements.

One can find a diffusion theory with quantum effects in some 
literatures. However, most of them handle the quantum diffusion 
process in a potential well. To the contrary, our problem is 
the quantum diffusion along a potential barrier. 
In order to adapt to this situation, i.e., to the diffusion process
along a potential barrier, especially at low temperatures,
we have developed a quantum diffusion theory that takes the 
quantum fluctuation due to the finite curvature of 
the potential barrier into account \cite{tawa04}. 
Our theory incorporates also a memory effect. 
In Ref. \cite{tawa04}, using a simplified model for the potential barrier,
mass and friction parameters, 
we reported that  
the quantum effects, especially memory effects, 
enhance the probability of overcoming the barrier
to form a compound nucleus 
compared with that calculated by 
assuming the standard fluctuation-dissipation 
theorem at low temperatures and for the potential curvature 
relevant to the synthesis of superheavy elements.  
In Ref. \cite{ayik05}, we developed a Langevin equation version of the 
quantum diffusion theory. Also, we reformulated so as to  
introduce the dissipation effect in a way more suitable to 
nuclear processes than the Caldeira-Leggett model 
adopted in Ref. \cite{tawa04}.

In this contribution we discuss quantum effects 
in the compound nucleus formation process
with more realistic parameters of the potential, mass, and friction 
and show that 
these effects enhance the compound nucleus formation probability 
at low temperatures
compared with that by the classical diffusion theory
using the standard fluctuation-dissipation relation.

\section{Quantum diffusion theory}
In this section, we explain two aspects of the quantum effects.

The first is that the connection between the diffusion 
and friction coefficients is modified from the 
well known fluctuation-dissipation theorem at high temperatures 
due to the quantum fluctuation originating from 
the finite barrier curvature.
For the diffusion process in a potential well, it   
is known that the ratio of the diffusion to the friction coefficients 
is given by
\begin{equation}
\frac{D}{\gamma}=\frac{1}{2}\hbar\Omega\coth(\frac{\hbar\Omega}{2T})
\label{fluc-diff-well}
\end{equation}
if the quantum fluctuation is taken into account. 
In Eq. (\ref{fluc-diff-well})
$D$, $\gamma$,
and $T$ are the diffusion and friction coefficients and 
the temperature, respectively. The $\Omega$ is defined by  
\begin{equation}
\Omega={\sqrt{\frac{V''(R_b)}{M}}},
\label{bcurvature}
\end{equation}
where $V''(R_b)$ and $M$ are the second derivative of the potential well 
at the bottom position of the potential $R_b$ and 
the mass parameter, respectively.
%
The relevant formula for the diffusion along a potential barrier 
such as the diffusion process around the conditional saddle point 
can be obtained by analytic continuation 
of Eq. (\ref{fluc-diff-well})
with respect to the frequency parameter $\Omega$
\cite{tawa04,hofmann98,rummel03}.
The result reads,
\begin{equation}
\frac{D}{\gamma}=\frac{1}{2}\hbar\Omega\cot(\frac{\hbar\Omega}{2T}),
\label{fluc-diff-barrier}
\end{equation}
with 
\begin{equation}
\Omega={\sqrt{\frac{\vert V''(R_B)\vert }{M}}},
\label{bcurvature-2}
\end{equation}
where $V''(R_B)$ 
is the second derivative of the potential 
at the barrier top position $R_B$.

One can easily confirm that 
both formulas, Eq. (\ref{fluc-diff-well}) 
and Eq. (\ref{fluc-diff-barrier}), reduce to 
the classical fluctuation-dissipation 
theorem, $D/\gamma=T$, in the high temperature limit where 
the thermal fluctuation far dominates the quantum fluctuation. 

\begin{figure}
  \includegraphics[width=.7\textwidth]{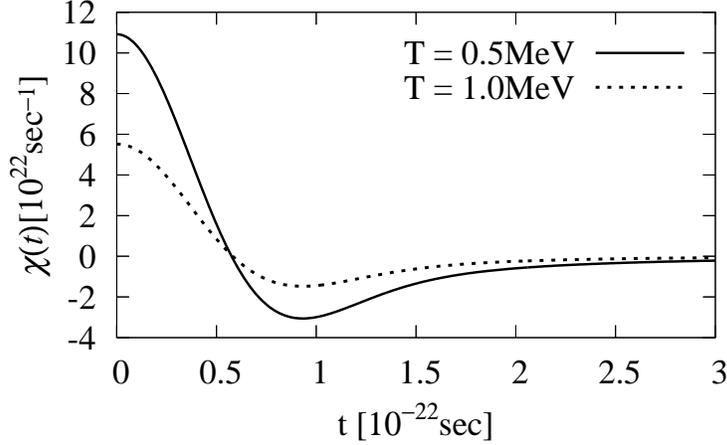}
  \caption{Correlation function as a function of time.
}
\end{figure}

The second quantum effect is the non-Markovian effect,
which leads to a colored noise problem.
The time correlation function
is given by \cite{tawa04,ayik05,cl83}
\begin{eqnarray}
  \langle R(t)R(t')\rangle &=&2\gamma T\cdot\chi(t-t')\label{randomforce},
\\
  \chi(t-t')&=&\int_{-\infty}^{+\infty}\frac{d\omega}{2\pi}
  e^{-i\omega (t-t')}\frac{\hbar\omega}{2T}\coth\frac{\hbar\omega}{2T}
  \cdot C(\omega),
\label{correlation}
\end{eqnarray}
to define the property of the random force $R(t)$.
%
The $C(\omega)$ is the cutoff function in the spectral density
of the heat bath which corresponds to the subspace of 
nuclear intrinsic degrees of freedom 
in heavy-ion collisions.
We employ the Gaussian form for the cutoff function, 
$C(\omega)=\exp\left[-(\hbar\omega)^{2}/2\Delta^{2}\right]$.
We can see from Eq. (\ref{correlation}) that the correlation function 
reduces to its classical Markovian form, 
$\chi(t-t')\rightarrow\delta(t-t')$, if the temperature is higher than 
the cutoff energy $\Delta$ and if the 
cutoff energy is sufficiently high.
To the contrary, at low temperatures, 
the quantum colored noise property of the random force 
needs to be seriously considered.

In Fig. 1, we show the correlation function as
a function of time for two temperatures, 
$T=0.5$ MeV (the solid line) and $1.0$ MeV (the dotted line). 
The cutoff energy has been fixed to be $\Delta=$ 15 MeV
as in Ref. \cite{ayik05}. At these temperatures, the 
non-Markovian effect is significant.

\section{Result}
In applying our theory to 
the compound nucleus formation process 
we use the liquid drop model \cite{krappe79} 
to calculate the potential energy surface in the space 
of nuclear deformation,
the hydrodynamical mass \cite{davis76} for the mass parameter, 
and the one-body dissipation \cite{blocki78} for the 
friction tenser.
The colored noise random force is handled by 
the spectral method given in Ref. \cite{spectral}.
We choose the $^{100}$Mo + $^{100}$Mo and $^{110}$Pd + $^{110}$Pd
systems, whose experimental fusion cross sections
are available, 
to examine the quantum effects on the probability
of overcoming the conditional saddle to form a compound nucleus. 
%
%
We use the separation distance between two fragments to 
describe the dynamics from inside the Coulomb barrier to 
inside the conditional saddle and determine 
its time evolution by 
solving the Langevin equation 
for a single macroscopic variable one hundred thousands times.
The other macroscopic degrees of freedom 
in the two center shell model parametrization \cite{maruhn72} 
are frozen during the compound nucleus formation process as; 
the mass partition parameter $\alpha=0$, the deformation parameter 
$\delta_1=\delta_2=0$, and the neck parameter $\epsilon=0.9$.
We initiate each trajectory from the touching configuration 
of the two fusing nuclei with zero momentum.
This corresponds to assuming that 
a strong energy dissipation from the macroscopic motion, i.e.,
the relative motion between the fusing two fragments, to 
nuclear intrinsic motions takes place inside 
the Coulomb barrier.
For simplicity, we ignore the change 
of the temperature of nuclear intrinsic degrees of freedom
during the time evolution of the system. 

\begin{figure}
  \includegraphics[width=1.\textwidth]{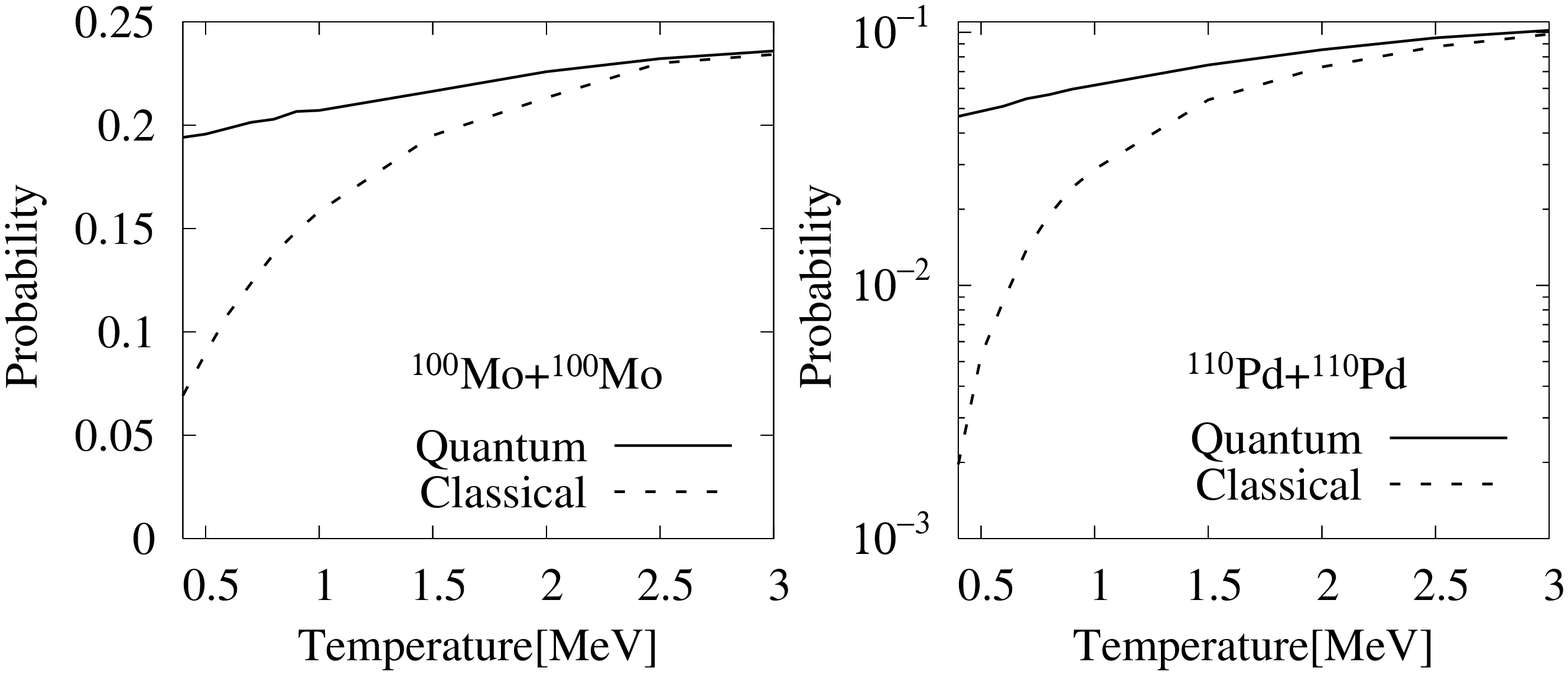}
\caption{
The compound nucleus formation probability as a function of 
nuclear temperature.
The systems are $^{100}$Mo +$^{100}$Mo (the left panel) 
and $^{110}$Pd +$^{110}$Pd (the right panel) reactions.
Note that different scales are used for the ordinates in 
the right and left figures.}
\end{figure}
 
Figure 2 compares the compound nucleus formation probability 
as a function of nuclear temperature
calculated by the quantum diffusion theory 
and by the classical diffusion theory
which postulates the standard fluctuation-dissipation relation
for the $^{100}$Mo +$^{100}$Mo (the left panel) 
and for the $^{110}$Pd +$^{110}$Pd (the right panel) reactions.
The solid lines are the results of the quantum diffusion theory,
while the dashed lines the classical diffusion theory.
These figures show that quantum effects become significant at 
low temperatures relevant to the experiments to 
synthesize superheavy elements.
They increase the compound nucleus formation probability 
at low temperatures.

\section{Summary}
We discussed quantum effects in the formation process
of a heavy compound nucleus described as a 
diffusion process along a potential barrier.
We have shown that 
the quantum effects increase the compound nucleus formation probability 
at low excitation energies, 
which are relevant to the synthesis of superheavy elements.

Further developments are needed 
in various aspects to apply the theory to more realistic problems. 
One of the essential developments is 
to generalize the present model which explicitly handles only 
one macroscopic variable, i.e., the relative distance between the 
colliding fragments,  
to the diffusion process in a multidimensional space by taking, 
for instance, the mass partition into account.
This will be crucial to discuss the competition between the
complete fusion and quasi-fission by including quantum effects.


\begin{theacknowledgments}
We wish to thank Prof. T. Wada and Dr. T. Asano for useful discussions.
Three of us (K.W., S.A., and N.T.) gratefully acknowledge the Physics
Department of the Middle East Technical University for the warm hospitality 
extended to us during our visit.
This work was supported in part by
the 21st Century Center of Excellence
Program ``Exploring New Science by Bridging Particle-Matter
Hierarchy'' of the Tohoku University.

\end{theacknowledgments}


\begin{thebibliography}{99}

\bibitem{swiatecki} W. J. Swiatecki, \emph{Physica Scripta},
\textbf{24}, 113--122 (1981).

\bibitem{aritomo99}Y. Aritomo, T. Wada, M. Ohta, and Y. Abe,
 \emph{Phys. Rev. C} \textbf{59}, 796--809 (1999). 


\bibitem{abe02} C. Shen, G. Kosenko and Y. Abe, 
\emph{Phys. Rev. C} \textbf{66}, 061602(R)1--5 (2002).

\bibitem{aritomo04} Y. Aritomo and M. Ohta, \emph{Nucl. Phys.} 
{\bf A744}, 3--14  (2004). 

\bibitem{tawa04} N. Takigawa, S. Ayik, K. Washiyama, and S. Kimura, 
\emph{Phys. Rev. C} \textbf{69}, 054605--1--5 (2004).

\bibitem{ayik05} S. Ayik, B. Yilmaz, A. Gokalp, O. Yilmaz, and N. Takigawa, 
\emph{Phys. Rev. C} \textbf{71}, 054611--1--8 (2005).

\bibitem{hofmann98} H. Hofmann, \emph{Phys. Rep.}
\textbf{284}, 137--380 (1998).

\bibitem{rummel03} C. Rummel and H. Hofmann, 
\emph{Nucl. Phys.} \textbf{A727}, 24--40 (2003).


\bibitem{cl83} A. O. Caldeira and A. J. Leggett, 
\emph{Physica} \textbf{121A}, 587--616 (1983).





\bibitem{krappe79} H. J. Krappe, J. R. Nix, and A. J. Sierk,
\emph{Phys. Rev. C} \textbf{20}, 992--1013 (1979).


\bibitem{davis76} K. T. R. Davies, A. J. Sierk, and J. R. Nix,
\emph{Phys. Rev. C} \textbf{13}, 2385--2412 (1976).


\bibitem{blocki78} J. Blocki, Y. Boneh, J.R. Nix, J. Randrup, 
M. Robel, A.J. Sierk and W.J. Swiatecki, 
\emph{Ann. Phys.} \textbf{113}, 330--386 (1978).

\bibitem{spectral} A. H. Romero and J. M. Sancho, 
\emph{J. Comp. Phys.}, \textbf{156}, 
1--11 (1999).



\bibitem{maruhn72} J. Maruhn and W. Greiner, 
\emph{Z. Phys.} \textbf{251}, 431--457 (1972).


\end{thebibliography}
\end{document}